\documentclass[
   prc,preprint,
  superscriptaddress,
  ]{revtex4}
\usepackage{graphicx}
\usepackage{multirow}

\usepackage[T1]{fontenc}
\usepackage[utf8]{inputenc}
\usepackage[english]{babel}
\usepackage{blindtext}
\usepackage{xcolor}
\begin{document}

\title{{Masses of fully-heavy tetraquark states from a four-quark
 static potential model }}
\author{{\bf Z. Asadi\footnote{E-mail: z.asadi@razi.ac.ir}}}
\affiliation{  Department of Physics, Razi University, Kermanshah 67149, Iran}
\author{{\bf G. R. Boroun\footnote{E-mail: boroun@razi.ac.ir}}}
\affiliation{  Department of Physics, Razi University, Kermanshah 67149, Iran}


\begin{abstract}
 In this article, we study the ground-state mass of full-heavy tetraquarks $cc\bar c\bar c$ and $bb\bar b \bar b$ with solving the nonrelativistic four-body systems. The flux-tube configurations, the tetraquark four-body potential butterfly, and the dimeson potential flip-flop of  SU(3) lattice quantum chromodynamics have been applied to describe the tetraquark interaction.
  Our numerical analysis indicates that the disconnected and connected static potentials can predict the mass of tetraquark very close to experimental data.
 

\end{abstract}

\maketitle

\section{INTRODUCTION}
Gell-Mann and Zweig had schematized the multiquark states with four or more quarks to interpret the observed spectrum of mesons and baryons products such as  $qq\bar q \bar q$ and $qqqq\bar q$~\cite{Gell,Zweig}.
 In a short time, enormous works have been done to explain the structure of four-quark states containing at least one light quark~\cite{Brambilla, Chen}. 
  The full-heavy tetraquark states have not been investigated until that; the very large energy states have been detected in the experiments of pair production of $\Upsilon(1s)$~\cite{Khachatryan}. The CMS Collaboration  observed pair production of $\Upsilon(1S)$ mesons in proton-proton collisions at $\sqrt{s}= 8$ TeV with a global significance of $3.6\sigma$ and a mass $18.4\pm 0.1(stat) \pm 0.2(syst)$GeV~\cite{Khachatryan,Durgut}. After that, the LHCb Collaboration studied the $\Upsilon(1S)_{\mu^+ \mu^-}$  invariant-mass distribution to seek a possible $bb \bar b \bar b$ exotic meson but they did not see any significant excess in the range $17.5-20.0$ GeV~\cite{Aaij}.  Another interesting structure is the full charm tetraquark. The existence of these particles has been confirmed by experimental searches of the  LHCb observation of the tetraquark containing only the charm quarks~\cite{Aaij2}. The LHCb Collaboration declared a narrow structure, matching the line shape of resonance and a broad structure next to the $di-J/ \Psi$  mass threshold with the data in the range of 6.2–7.2 GeV~\cite{Aaij2}, which called X(6900) with the structure $cc \bar c \bar c$. The existence and stability of such states have been considered by applying several models with different interactions~\cite{Berezhnoy,Maiani3,Wu,Bai,Hughes,Buccella,Karliner,Maiani2}.
 

Creutz  applied the lattice QCD simulations with the Wilson loop  to describe the interquark potential between a
   quark and an antiquark~\cite{Creutz} after that 
  a large amount of effort has been devoted in lattice QCD to 
  study the multiquark force~\cite{Creutz, Bali, Takahashi, Alexandrou, Okiharu}. These potentials are successful  to calculate the energy  and the mass of tetraquark systems contained purely heavy
  quarks~\cite{Csikor,Chiu,Sasaki}.
 The experimental discoveries of multiquark hadrons 
  reveal new aspects of  the interquark force such as the quark confinement force, the
 color-magnetic interaction and the diquark correlation~\cite{Jaffe}.
 According to these,  the proper Hamiltonian for
 the quark-model calculation of multiquarks has been suggested to
  investigate the interquark force in the multiquark system directly based on QCD ~\cite{ Stancu,Kanada-Enyo}.
  
 A number of phenomenological models have
  been put forward to explain multiquark stability.  A type-II diquark-antidiquark model, proposed by L. Maian et al. based on the new Ansatz on spin-spin couplings, in which the cq (c
  the charm and q a light quark) interaction inside the diquark dominates over all other possible pairings~\cite{Maiani1}. In this model, diquarks are similar to
  compact bosonic building blocks which supposed that the size of the entire tetraquark is consistently larger than the size of these blocks therefore, the spin-spin interactions between different diquarks are neglected. The color configuration for the diquark-antidiquark pair is similar to one of the quark-antiquark systems with a bound state of two pointlike color sources~\cite{Maiani1}. The dynamical diquark method was introduced to explain the nature of the exotic XYZ states based
     on a  color flux-tub configuration where the separated diquark-antidiquark pair is connected by the shortest tube configurations (energetically),~\cite{Lebed2,Lebed5,Lebed3,Lebed4}. The flux tube has different configurations, which we will discuss in the next section, but none of these structures can be suitable to describe the dynamical diquark model especially as the system remains
     in a state of rapid change until the moment it decays~\cite{Lebed5}. In this physical
  picture, each diquark is only in the color-triplet
     combination~\cite{Lebed}. In 2020, Lebed et al. investigated the basic spectroscopy of $c \bar c  c\bar c$ tetraquark states in this model~\cite{Lebed}.
    The full mass spectrum of all tetraquark states was predicted, using lattice-calculated confining two-body potential computed with the
  Born-Oppenheimer (BO) approximation, which includes spin and isospin-dependent operators with the spin-orbit and tensor coupling operators, such a way that the  spin-spin couplings dominated between quarks (antiquarks), not between quark and
    antiquark~\cite{Lebed}.
   
  The flux-tube picture as an analytical model for the multiquark system  supported by  lattice QCD has been considered for the structure and the reaction of
  hadrons~\cite{Okiharu,Alexandrou,Nambu,Kogut,Carlson,Isgur,Suzuki}.
   The evaluated tetraquark potentials in Refs.~\cite{Okiharu, Alexandrou} have been constructed the Wilson loops for investigating the interaction between quarks in the four-quark system directly from QCD by using SU(3) lattice QCD at the quenched level and studied the hypothetical flux-tube pictures for the multiquark system. They [i.e, Refs.~\cite{Okiharu, Alexandrou}] used three flux-tube configurations according to four quarks' locations. All quarks and antiquarks are connected with the single flux-tube in the connected flux tube system while in the  “dimeson” states there are two disconnected flux-tubes~\cite{Okiharu}. The investigating of the properties of tetraquark using these three types of four-quark potentials and the transition between the connected and disconnected four quark states are physically important for analyzing the tetraquark decay process into two mesons and the reaction mechanism between two mesons. 
      In this paper,  we have calculated masses of full-heavy tetraquarks $cc\bar c\bar c$ and $bb\bar b \bar b$ using three types of connected and two-meson four-quark potentials.
   
  
   This paper is organized as follows. Section.~\ref{S2} gives a brief review on the color flux-tube model. The calculation method is described in Sec.~\ref{S3}. The numerical results for both the spin-independent  and spin-dependent potential  are presented in Sec.~\ref{S4}.
  Section ~\ref{RandDD} is devoted for summary and concluding remarks.
%

\section{THE COLOR FLUX-TUBE MODEL}\label{S2}

The short-distance  one gluon exchange (OGE) Coulomb force as a quantity of  perturbative QCD and the long-distance
 confinement force as a typical contribution of nonperturbative can be developed to an alternative color flux tube based on the lattice QCD picture with the Wilson loops~\cite{Takahashi,Alexandrou,Okiharu}.
 The tetraquark SU(3) Wilson loop is defined in a four quarks  gauge,
       at time $t =0$ which annihilated at a later time $t$. The potential of the system has been extracted from the behavior of the Wilson loop at the large time region~\cite{Alexandrou}. 
The multibody color flux-tube dynamical mechanism has been used to describe multiquark
states from this phenomenological point of view~\cite{Deng,Deng2}.
 
\begin{figure}
 	
 	\resizebox{0.6\textwidth}{!}{%
 		\includegraphics{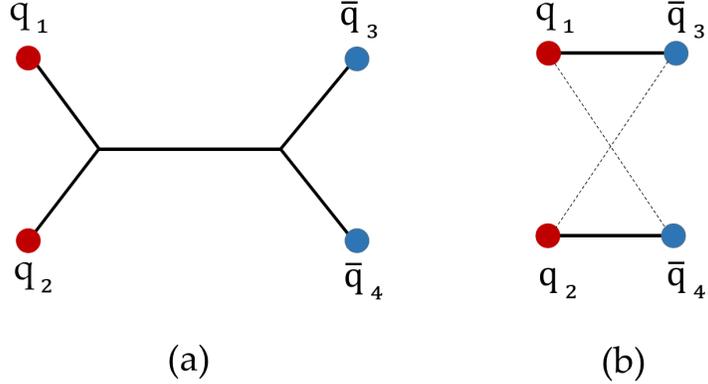}}
 	\caption{(a)The connected tetraquark system, which is the butterfly configuration.(b)The disconnected tetraquark system, which corresponds to the flip-flop configuration.}
 	\label{11}
 \end{figure}

The minimum energy of a system of two quarks and two antiquarks is recorded when the two quarks and the
two antiquarks are linked by the minimal value of the total flux tube length. Therefore the flux tube is formed to achieve a minimum of the total flux-tube length of the system
for the low-lying state~\cite{Okiharu}. The theoretical form of the tetraquark potential according to the four quarks location has three candidates for the flux-tube configuration. One for the four-quark state of two quarks and two antiquarks $(qq\bar q \bar q)$ that fluxtube system connect all quarks and antiquarks with the single flux tube as shown in the (a) panel of Fig.~\ref{11}, so-called “butterfly” configuration. Two others for the tetraquark states of two mesons $(q\bar{q} q \bar{q})$  with two disconnected flux tubes as shown in the (b) panel of Fig.~\ref{11}, the “flip-flop”
configurations. The theoretical form of
the tetraquark potential for each configuration is written as below.

The  OGE plus the double-Y-shaped flux-tube configurations have been suggested for the theoretical form of the tetraquark $(qq \bar q \bar q )$ potential.  
All the quarks and antiquarks are in the same plane and link to each other with the flux tubes that the angles between them must be $120^0$ \textcolor{blue}~\cite{Alexandrou}. 
The tetraquark potential for the butterfly is written as 
\begin{eqnarray}
\label{eqn0}
V_{2q 2\bar q}^{Butterfly}& =& - A_{4q}[\frac{1}{2}(\frac{1}{r_{13}}+\frac{1}{r_{14}}+\frac{1}{r_{23}}+\frac{1}{r_{24}})
+(\frac{1}{r_{12}}+\frac{1}{r_{34}})] + \sigma_{4q} L_{min}  
\end{eqnarray}
with $A_{4q}$ the Coulomb coefficient,  $\sigma_{4q}$ the string tension extracted
from the quark-antiquark potential and $r_{ij}=|r_j-r_i|$ with $r_i$ which denotes the location
of $i$th particle.  $L_{min}$ is the minimal value of the total flux-tube length in the (a) panel of Fig.~\ref{11}.
The minimal length for the tetraquark system accrues when the flux tubes from each of the quarks and the antiquarks meet at two Steiner points [Steiner point is the junctions of two flux tubes $(s_1 , s_2) $]~\cite{Warren}. $L_{min}$ can be  expressed as
\begin{eqnarray}
\label{eqn2}
L_{min}= r_{1s_1} + r_{2s_1} + r_{3s_2} + r_{4s_2} + r_{s_1s_2}.
\end{eqnarray}

A disconnected flux-tube configuration the (b) panel of Fig.~\ref{11}
is acceptable when the nearest quark and antiquark pair
is spatially close. Thus the system can be regarded as a
“dimeson state” rather than a connected four quark state. The  
 flip-flop configuration is  summarized in the form
\begin{eqnarray}
\label{eqn3}
V_{2(q \bar q)}^{ flip-flop} &=& -A_{q\bar q}(\frac{1}{r_{13}}+\frac{1}{r_{24}})+\sigma_{q \bar q} (r_{13}+r_{24}).
\end{eqnarray} 
Where the potential for dimeson system the (b) panel of Fig.~\ref{11}  is given by exchanging indices  $1\leftrightarrow 2$.
The  total tetraquark potential is defined to  represent the minimum energy of the connected four quarks; it is taken by 
\begin{eqnarray}
\label{eqn4}
V_{4q}& =& min (V_{2q 2\bar q}^{Butterfly},V_{2(q \bar q)}^{ flip-flop}).
\end{eqnarray}
The transition between the connected and disconnected four-quark states can be demonstrated by Eq.~(\ref{eqn4}) where these two potentials are equal.

   The lattice QCD simulation parameters such as $\beta$, the lattice size, and the corresponding lattice
   spacing $a$   with some related information are estimated so as to reproduce the string tension $\sigma$.  
    In Table.~\ref{tab10} the lattice parameters, Coulomb coefficient, the string tension $\sigma$ and the quark mass  $m_c$ and   $m_b$ have been presented. They have been obtained  by using the
   fitting analysis on the on axis data of the quark-antiquark potential in lattice QCD ~\cite{Okiharu, Takahashi2}. The extensive studies on the interquark potentials in lattice QCD    have  indicated  that  $ \sigma_{4q} \simeq \sigma_{q \bar q}$ and the OGE results for the  Coulomb
       coefficient $A_{q \bar q} \simeq 2A_{4q} $ are fairly compatible  with the hypothetical flux-tube picture~\cite{Okiharu}.
    \begin{table*}
       	\caption{{\label{tab10}  The quark mass for each $\beta$, the lattice size, and the lattice spacing a with the  Coulomb coefficient and string tension.
       	}}
       	
       	\begin{tabular}{c@{\hskip 0.15in}c @{\hskip 0.15in}  c@{\hskip 0.15in}c@{\hskip 0.15in}c@{\hskip 0.15in}c@{\hskip 0.15in}c@{\hskip 0.15in}c}\hline
       		$\beta $~\cite{Okiharu}& Lattice size~\cite{Okiharu} &  a(fm)~\cite{Okiharu} & $\sigma_{q \bar q}(GeV)^2$~\cite{Okiharu} &$A_{q \bar q}$~\cite{Okiharu}&$m_{c}(GeV)$ & $m_{b}(GeV)$& S.N \\ \hline \hline
       		$6.0$&$16^3\times 32$&0.10& 0.09&0.27&1.70&4.61& I \\ 
       		$5.8$&$16^3\times 32$ &0.14&0.18&0.27& 1.88&4.79& II  \\
       		 [1ex] \hline

       	\end{tabular}\label{tab10}
       \end{table*}
    
     For the four-quark system, the Hamiltonian  can be expressed as follows:
       \begin{eqnarray}
           \label{eqn70}
            H= \sum_{i=1}^4{(m_i+\frac{{p_i}^2}{2m_i})}-T_{cm}+V_{4q},
          \end{eqnarray} 
          $m_i$ and $P_i$ are , respectively the mass and momentum of the ith quark (antiquark). $T_{cm}$ is the center-of-mass kinetic energy
          of the states. The solution of this four-body problem is very difficult.
         We partially reduce the difficulty of the calculation with the  some simple assumptions. The quarks are assumed to be heavy quarks and nonrelativistic. All of them, quarks and antiquarks, have the same mass.     
          
   The model parameters in Table.~\ref{tab10} have been adopted to calculate the mass of experimentally observed full charm tetraquark X(6900) and predict the mass of the full heavy bottom tetraquark.
   
    
   

   \section{Numerical Calculation Method }\label{S3} 
   The wave function of tetraquark can be constructed of the spatial degrees of freedom and the internal degrees of freedom of spin, color, and flavor.
       Tetraquarks have been comprised of two couples of identical fermions; thus, their wave functions must be antisymmetric for the exchange of the two quarks and the two antiquarks, and like all physical states, they are the color singlet. The trial wave function of the tetraquark can be defined as
   
     \begin{eqnarray} 
     \label{eqn60}    
        \Psi^{q_1q_2\bar q_3 \bar q_4}_{JM_J}=  \mathcal {A} \{[[[\phi_{nlm}(\rho_{12})\chi_s(12) ]_{\Lambda M_{\Lambda}}^{q_1q_2}[\phi_{NLM}(\rho_{34})\chi_{s'}(34)]_{IM_I}^{\bar q_3 \bar q_4 }]^{q_1q_2\bar q_3 \bar q_4}_{\lambda M_{\lambda}}\\ \nonumber
        \times \phi_{\nu \lambda \mu}(r_{12-34})]_{JM_J}^{q_1q_2\bar q_3 \bar q_4}
        \times [\eta_c(12)\eta_c(34)]_{CW_c} \times [\xi_f(12)\xi_f(34)]_{FW_f} \}.
       \end{eqnarray}
       
    The factor $\mathcal {A}$ is  the antisymmetrization operator and $\phi$, $\chi$, $\eta$, and $\xi$  respectively, express space, spin, color, and flavor states. All the indexes are all possible flavor–spin–color–spatial intermediate quantum numbers.

   To use the planar geometry, it has been supposed all the quarks are in one plate.
   Thus, reducing the number of variables of the four body potentials is useful to simplify the calculation method. The new relative spatial coordinates can be defined as
   \begin{eqnarray}
      \label{eqn9}
        \rho_{12} &=& r_1 -r_2 \\
          \rho_{34} &=& r_3 -r_4 \\
          r_{12,34} &=& \frac{m_1r_1+ m_2r_2}{m_1+m_2} -  \frac{m_3r_3+ m_4r_4}{m_3+m_4}\\
          R &=&  \frac{m_1r_1+ m_2r_2+m_3r_3+ m_4r_4}{m_1+m_2+m_3+m_4}        
      \end{eqnarray}
   In our model, the interdistance of two quarks with the interdistance of two antiquarks are same, $ \rho_{12}= \rho_{34}= \rho$. With this simplification,  only two dimensions remain in Eq.~(\ref{eqn70})  r the inner distance between quark and antiquark and $\rho$ the distance between two quarks and two antiquarks. The excitations  orbital
   angular momenta associated with the Jacobi coordinates  are $l_{\rho}$ and $l_{r}$. The parity of the tetraquark system can be expressed in terms of the relative orbital angular momentum as $P=(-1)^{l_{\rho}+l_{r}}$. To calculate the ground state of energies, the system has been considered in positive parity with $l_{\rho} = 0$ and $l_{r} = 0$.
   
\begin{figure}
    	
    	\resizebox{0.3\textwidth}{!}{%
    		\includegraphics{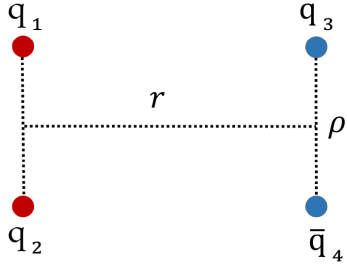}}
    	\caption{A planar configuration of the tetraquark system.}
    	\label{12}
    \end{figure}

    Based on the planer geometry of tetraquark in Fig.~\ref{12}, the   butterfly and flip-flop potentials can simplify as follows:
    
   \begin{eqnarray}
   \label{eqn0}
   V_{2q 2\bar q}^{Butterfly}& =& - A_{4q}[\frac{1}{r}+\frac{1}{\sqrt{r^2+\rho^2}}
   +\frac{2}{\rho}] + \sigma_{4q} (r+\sqrt{3}\rho)  
   \end{eqnarray}
   
   \begin{eqnarray}
   \label{eqn3}
   V_{2(q \bar q)}^{ flip-flop} &=& -A_{q\bar q}\frac{2}{r}+\sigma_{q \bar q} (2r).
   \end{eqnarray} 
    
    The total tetraquark potential in this two-dimensional system has been defined in Eq.~(\ref{eqn4}).
    Finally, the Schrodinger equation corresponding to two relative
    motions $\rho$ and $r$ has been defined as:
     \begin{eqnarray}
       \label{eqn130}
    [\sum_{i=1}^{4}m_i-\frac{\hbar}{2m}(\nabla ^2_{\rho}+\nabla ^2_r)-T_{cm}+V_{2(q \bar q)}^{ flip-flop}(r, \rho)]\Psi(r,\rho)= E \Psi(r,\rho)
       \end{eqnarray} 
       This equation is somewhat complicated to get a reliable result; thus, 
        a high-precision numerical method is required. The Gaussian expansion method (GEM)~\cite{Hiyama} has been used to study this four-quark system. The GEM was proposed to a  variety of few-body systems and has been powerful to solve three and four-body problems. 
              In this method, the spatial wave function has been expanded in  terms of a set of Gaussian basis functions,
       \begin{eqnarray}
              \label{eqn140}
           \Psi_{lm}^G (R)= \sum_{n=1}^{n_{max}}c_{nl}N_{nl}R^l e^{-\nu_nR^2} Y_{lm}(R)\\
           N_nl = (\frac{2^{l+2(2\nu_n)^{(l+\frac{3}{2})}}}{\sqrt{\pi}(2l+1)!!})^{\frac{1}{2}} \quad \textrm{} \quad   \\
           \nu_n = \frac{1}{R^2_n}\quad  a = (\frac{R_{n_{max}}}{R_1})^{\frac{1}{n_{max}-1}}   \quad . 
        \end{eqnarray} 
       
       There are three parameters $\{n_{max},R_1,R_{n_{max}}\}$ that we  have calculated the numerical result with $n_{max}= 10$, $R_1= 0.1 fm$ and $R_{n_{max}} = 3.0$.
       The spin wave functions base on total spin of the tetraquark   $S_{tot}=0,1,2$ in the ($qq)(\bar{q}\bar{q}$) configuration, are in the  six below states.
           \begin{eqnarray}
           \label{eqn10}
               \chi_{s=0}(12,34) =\left\{ 
                                  \begin{array}{ll}
                                     |\chi_{s=0}(12)\chi_{s=0}(34)>\\ |\chi_{s=1}(12)\chi_{s=1}(34)>
                                     
                                   \end{array}
                                    \right.
                                                            \\
                    \chi_{s=1}(12,34) =\left\{ 
                    \begin{array}{ll}
                      |\chi_{s=0}(12)\chi_{s=1}(34)>\\
                       |\chi_{s=1}(12)\chi_{s=0}(34)>                                                             
                         \end{array}
                           \right.               \\
                     \chi_{s=2}(12,34) =
                   |\chi_{s=1}(12)\chi_{s=1}(34)>                                                            
                     \end{eqnarray}
       
   The color representation for the tetraquark is only one single and also  the configuration of diquark and  diantiquark are antisymmetric $[q_1q_2]_{\bar 3_c}$ and  $[\bar q_1 \bar q_2]_{3_c}$ and symmetric $[q_1q_2]_{6_c}$ and  $[\bar q_1 \bar q_2]_{\bar6_c}$~\cite{Deng}. Therefore, the allowed color tetraquark $[qq][\bar q \bar q]$ state is in the two representations $[3]_c\bigotimes[\bar3]_c$ and $[6]_c\bigotimes[\bar 6]_c$. 
   
    \section{Numerical results and analysis }\label{S4} 
    \subsection {Spin-independent potential}
\begin{figure}
                     	
                     	\resizebox{0.5\textwidth}{!}{%
                     		\includegraphics{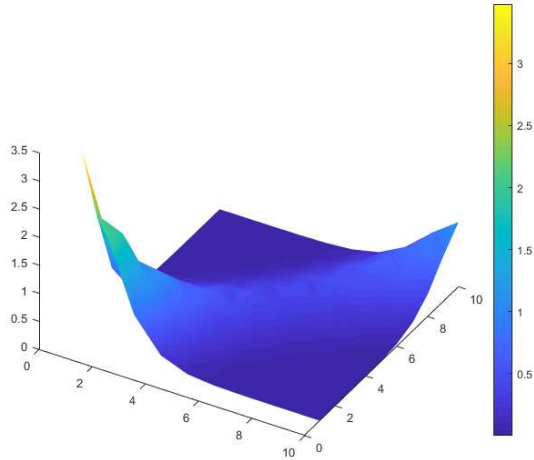}}
                     	\caption{The visualization of ground state energy for the flip-flop potential with ignoring the OGE part}
                     	\label{111}
                     \end{figure}

\begin{figure}
                                          	
                	\resizebox{0.5\textwidth}{!}{%
              	\includegraphics{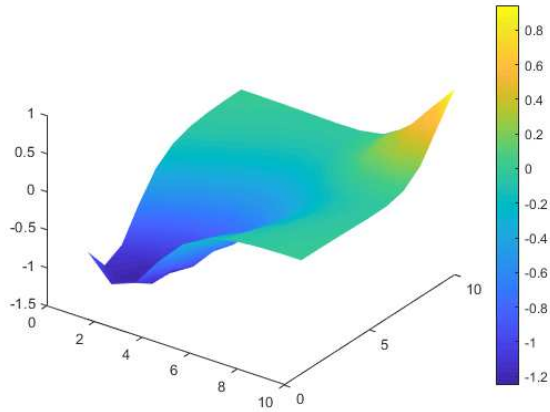}}
        	\caption{The visualization of ground state energy for the flip-flop potential}
                 	\label{112}
                \end{figure}

    We consider solving the two-dimensional Schrodinger equation, which is now equivalent to a matrix eigenvalue equation of dimension $n_rn_{\rho}\times n_rn_{\rho}$, using the Rayleigh–Ritz variational method,
\begin{eqnarray}
   \label{eqn80}
       <\Psi_{lm,n}^G |H-M|\Psi_{lm,n^\prime}^G > =0 .
    \end{eqnarray}
    Here, M is the mass of tetraquark, which has been calculated by applying the parameters of Table.~\ref{tab10}.  The visualization of the ground state energy for the flip-flop potential with ignoring the OGE part is presented in Fig.~\ref{111}. Adding the OGE part changes the visualization which is depicted in Fig.~\ref{112}. As one can see, the matrix elements of the ground state without the OGE part are positive with a big peak at the first element, which is caused by the arrangement of  Gaussian size parameters. After adding the OGE part to the calculation the elements of the matrix  started to be negative, which means the contribution of the OGE part leads to the achievement of the bound state energy points, which are visible with small pits in Fig.~\ref{112}. These points help to produce the eigenvalues that reported the best value of the tetraquark mass with different static potentials.
          
         
         The $cc \bar c \bar c $ and $bb \bar b \bar b $ masses  without a spin-dependent correction with two sets of parameters Table.~\ref{tab10} for different static potentials  have been presented in Table.~\ref{tab101}. In this table, the calculated masses compared with the experimental mass of the X(6900) for $cc \bar c \bar c$ and the observed mass for pair production of $\Upsilon(1S)$ mesons for $bb \bar b \bar b$. As one can see, the flip-flop potential result in S.N $I$ for $cc \bar c \bar c$ is $6.8878$ GeV and has a considerable agreement with the experimental mass of the X(6900). The obtained result for the butterfly is not significant as the flip-flop one but this is a  considerable result with just 0.04 GeV different from experimental data. Our results show that the flip-flop and butterfly potentials have almost the same results for the heavier tetraquark. The masses of $bb \bar b \bar b$,  which have been measured with these two potentials in S.N $I$ presented in Table.~\ref{tab101}, are in good agreement with the experimental mass of pair  $\Upsilon(1S)$ mesons.
         
          \begin{table*}
                    	\caption{{\label{tab101}  The ground state  masses of the fully heavy tetraquarks,  unit in GeV .
                    	}}
                    	
                    	\begin{tabular}{c@{\hskip 0.15in}c @{\hskip 0.15in}  c@{\hskip 0.15in}c@{\hskip 0.15in}c@{\hskip 0.15in}c@{\hskip 0.15in}c@{\hskip 0.15in}c}\hline
                    		System & M($flip-flop$)  & M($butterfly$)  & M($flip-flop+butterfly$)& S.N &$M_{exp}$\\  \hline
                    		\multirow{2}{*}{$cc \bar c \bar c$}&6.8878 &6.9456&6.8218& I & \multirow{2}{*}{$6.905 \pm 11$~\cite{Aaij2}} \\ \cline{2-5}
                    		  	&7.559&7.701& 7.694&II \\ \hline
                    		
                    	\multirow{2}{*}{	$bb \bar b \bar b$}&18.4397 &18.4441&18.4404& I& \multirow{2}{*}{$18.4\pm 0.1\pm.2$~\cite{Durgut}} \\ \cline{2-5}
                    	&19.186&19.302& 19.295& II \\ \hline

                    	\end{tabular}\label{tab101}
                    \end{table*}

\begin{figure}
                        	
                        	\resizebox{0.7\textwidth}{!}{%
                        		\includegraphics{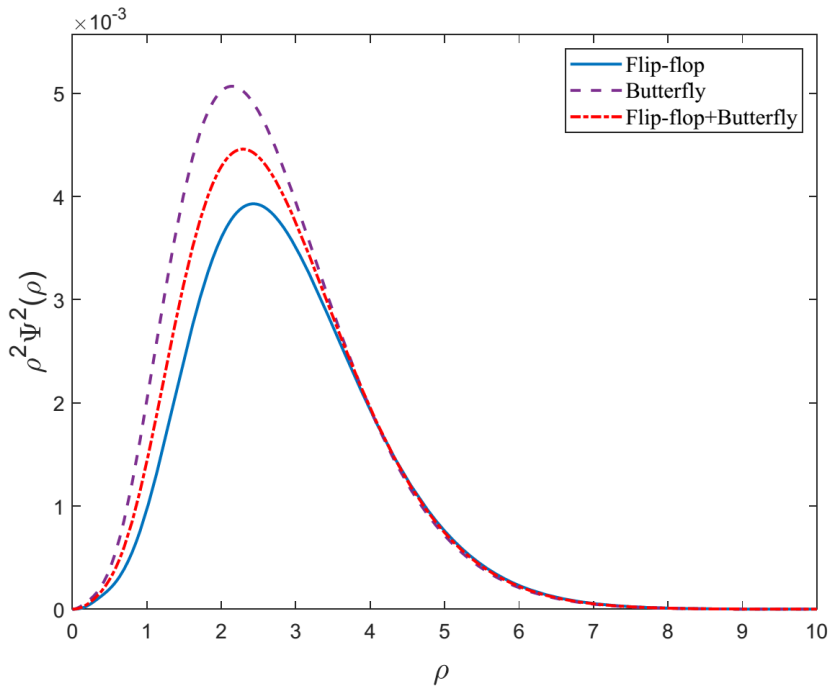}}
                        	\caption{ The radial probability density distributions of the initial state wave functions  for different static potentials using S.N $I$}
                        	\label{44}
                        \end{figure}

           
           The radial probability density distributions of the initial state wave functions for different static potentials using S.N $I$ have been depicted in Fig.~\ref{44}. The two-dimensional matrix eigenvectors of the tetraquark $cc \bar c \bar c$ have been used to generate the wave functions.  We can see that curves are localized in the range 2 to 3 $fm$ with an exponential tail at large distances. The maximum probability for the butterfly is located at $r=2.14$ $fm$. The location of maxima for the flip-flop  shift towards higher $r $ values at $2.42$ $fm$. The butterfly potential is localized sooner and stronger than the flip-flop potential, its squared wave function is around $22.5\%$ localized larger than the flip-flop one.
             
    \subsection {Spin-dependent corrections}
    The general form of the total spin-dependent interaction in QCD between the particles i and j for the four-body system  can be expressed as follows~\cite{Rujula}:
    \begin{eqnarray}
               \label{eqn122}
                  H^{SS} = \sum_{i<j=1}^{4}-\frac{2\pi \alpha_s \lambda_i . \lambda_i \sigma_i . \sigma_j \delta (r_{ij}) }{12 m_i m_j}.
                \end{eqnarray}
    Here $\alpha_s$ is the quark-gluon coupling constant and $\lambda$ and $\sigma$ represent the Gell-Mann matrices and Pauli matrices, respectively. The values of $<\lambda_i . \lambda_i>$  and $<\sigma_i . \sigma_j>$  have been measured  according to the symmetry properties of the tetraquark wave function~\cite{Deng3}. A tetraquark state with total quantum numbers $l = 0$, and $S = 0$, $ 1$ and $2$ with  positive parity  for  two representations $[\bar3]_c\bigotimes[3]_c$ and $[6]_c\bigotimes[\bar 6]_c$ has been considered.  The spin-dependent  correction with calculating the elements matrix $<\Psi| H^{ss}|\Psi>$ for three wave functions with total spin  $S = 0,$ $ 1$ and $2$ has been added to the spin-independent matrix. The masses of  states $cc \bar c \bar c$ and $bb \bar b \bar b$ for two color configurations with the spin-dependent correction are presented in Table.~\ref{tab102}.  It is worth highlighting that from Table.~\ref{tab102} the color configuration has a significant role in the effect of spin correction in the mass of tetraquark. But in our calculation for heaver tetraquark, this effect has almost vanished. As one can see in Table.~\ref{tab102} the mass of $cc \bar c \bar c $ for different spin quantum numbers have considerable change while no change has been measured for the mass of $bb \bar b \bar b$. There are no exact experimental observations for the mass of  X(6900) with different spin quantum numbers; nevertheless, the obtained masses with $J^P = 2^+$ for flip-flop and $J^P = 1^+$ for butterfly are approximately the same with the mass of X(6900).
    
    \begin{table*}
               	\caption{{\label{tab102}  The mass spectra of the ground state $cc \bar c \bar c $ and $bb \bar b \bar b$ for different static potentials, unit in GeV.
               	}}
               	
               	\begin{tabular}{c@{\hskip 0.15in}c @{\hskip 0.15in}  c@{\hskip 0.15in}c@{\hskip 0.15in}c@{\hskip 0.15in}c@{\hskip 0.15in}c@{\hskip 0.15in}c}\hline
               		System & Color & $J^P$  & $M(flip-flop) $ & $M(butterfly)$ &$M(flip-flop+ butterfly)$ \\  \hline
               	\multirow{6}{*}	{$cc \bar c \bar c$}&\multirow{3}{*}	{$[\bar3]_c\bigotimes[3]_c$}&$0^+$ &6.8500&6.8742&6.8217  \\ \cline{3-6}
               		  &&$1^+$& 6.8704&6.9133&6.8218 \\  \cline{3-6}
               		&&$2^+$& 6.9127&6.9899&6.8218 \\  \cline{2-6}
               		&\multirow{3}{*}	{$[6]_c\bigotimes[\bar 6]_c$}&$0^+$ &6.9285&7.0174&6.8218  \\ \cline{3-6}
               		               	&	&$1^+$&6.9660& 6.8992&6.8217 \\  \cline{3-6}
               		               	&	&$2^+$&6.8783&6.9281&6.8218 \\  \hline
               		\multirow{6}{*}	{$bb \bar b \bar b$}&\multirow{3}{*}	{$[\bar3]_c\bigotimes[3]_c$}&$0^+$ &18.43971&18.44416&18.44045  \\ \cline{3-6}
               		       		&&$1^+$&18.43971&18.44417&18.44045 \\  \cline{3-6}
               		       	&	&$2^+$&18.43972&18.44417&18.44045 \\  \cline{2-6}
           &\multirow{3}{*}	{$[6]_c\bigotimes[\bar 6]_c$}&$0^+$ &18.43972&18.44417&18.44045  \\ \cline{3-6}
             	&	&$1^+$&18.43971& 18.44417&18.44045 \\ \cline{3-6}
            	&	&$2^+$&18.43971& 18.44416&18.44045 \\  \hline    		       		
             	\end{tabular}\label{tab102}
               \end{table*} 
 In fact, there are many theoretical calculations for measuring the mass of fully-heavy tetraquark states which predict similar or different mass spectrums for $J^P = 0^+$, $1^+$ and $2^+$ corresponding to their method~\cite{Wang2,yang1}. The full mass spectrum of all tetraquark states was predicted in Ref.~\cite{Lebed} a state or
   states within the 2S multiplet were suggested as the best interpretation for identifying $X(6900)$~\cite{Lebed}. 
 The masses of the ground state of tetraquark $cc \bar c \bar c$  for different  $J^P$ quantum numbers have been compared with the other nonrelativistic quark models in Table.~\ref{tab104}. Our results are compatible with the other approaches; however, they are larger than the average of the other works. It has proved that  four-body interaction models are completely suitable to apply for studying the tetraquark properties as well as the other models.
  \begin{table*}
       	\caption{{\label{tab104}  The ground state  masses of $cc \bar c \bar c$ in various models,  unit in GeV.
                      	}}
                      	
        	\begin{tabular}{c@{\hskip 0.15in}c @{\hskip 0.15in}  c@{\hskip 0.15in}c@{\hskip 0.15in}c@{\hskip 0.15in}c@{\hskip 0.15in}c@{\hskip 0.15in}c}\hline
          	$J^P$& M($flip-flop$)  & M($butterfly$)  & M\cite{Chen2}& M\cite{Wu}&M\cite{Deng} &M\cite{Lloyd}&M\cite{Ader}\\  \hline
           	$0^+$ &6.8500&6.8742& $6.82\pm0.18$ & 7.016&6.491& 6.477&6.437  \\ \hline
              	$1^+$&6.8704&6.9133&$6.51\pm0.15$&6.899&6.580&6.528&6.437 \\ \hline
             	$2^+$&6.9127&6.9899&$6.51\pm0.15$&6.956&6.607&6.573&6.437 \\ \hline	
                                                     		
    	\end{tabular}\label{tab104}
        \end{table*}               
    In Table.~\ref{tab105} we have compared the  $J^P = 0^+$ , $1^+$, and $2^+$  ground state masses of tetraquark $bb \bar b \bar b$  with the works  that reported results on $bb \bar b \bar b$
         using different theoretical techniques. It would be noted that in many works authors used calculation methods~\cite{Wu,Deng,Deng2,Deng3} and physical pictures~\cite{Gordillo,Lebed4,Maiani2,Maiani3} which are so close to our method but the potentials we have applied to describe the connection between four quarks are different. We believe that the main reason for the slight difference between our results and other works is the difference in potentials.
      \begin{table*}
            	\caption{{\label{tab105}  The ground state  masses of $bb \bar b \bar b$ in various models,  unit in GeV.
                           	}}
                           	
             	\begin{tabular}{c@{\hskip 0.15in}c @{\hskip 0.15in}  c@{\hskip 0.15in}c@{\hskip 0.15in}c@{\hskip 0.15in}c@{\hskip 0.15in}c@{\hskip 0.15in}c}\hline
               	$J^P$& M($flip-flop$)  & M($butterfly$)  & M\cite{Chen2}& M\cite{Wang}&M\cite{Wu} &M\cite{Liu}&M\cite{Gordillo}\\  \hline
                	$0^+$ &18.43972&18.44417& $18.45 \pm 0.15$ & 18.840&20.155& 19.322&19.199  \\ \hline
                   	$1^+$&18.43972&18.44417&$18.33 \pm0.17$&18.840& 20.212&19.329&19.276 \\ \hline
                  	$2^+$&18.43972&18.44417&$ 18.32\pm 0.17$&18.850& 20.243&19.341&19.289 \\ \hline	
                                                          		
         	\end{tabular}\label{tab105}
             \end{table*}

\section{CONCLUSION }\label{RandDD}
Masses of fully-heavy tetraquark for different flux-tube configurations, the connected butterfly, and the disconnected flip-flop configuration have been computed with solving the nonrelativistic four-body systems. The measured mass of $cc \bar c \bar c$ with the flip-flop potential is $6.8878$ GeV which is so close to the experimental mass of X(6900) while for butterfly configuration prediction is about $6.9456$ Gev that is about $0.046$ GeV above the experimental data. After adding spin-dependent correction in color configuration $[3]_c\bigotimes[\bar 3]_c$ with total spin 2 the flip-flop potential has reported the best result for the mass of $cc \bar c \bar c$, it is $6.9127$ GeV. The mass of $bb \bar b \bar b$   measured with the flip-flop is about $18.440$ GeV with S.N I  which is in good agreement with the experimental mass of pair  $\Upsilon(1S)$ mesons. There was a perceived no significant change for the masses of the $bb \bar b \bar b$  tetraquark with the different spin quantum numbers after adding the spin-dependent correction. Altogether both potentials provide good results for the mass of tetraquarks. However, the results have shown that disconnected configuration is a little more suitable structure for the heavy tetraquark. The comparison between our results and the works which used the two-body potential proves that the four-body potentials are also reliable for studying the properties of tetraquark.  

%


  %


\acknowledgements
{ The authors are thankful to the Razi University for
	financial support of this project. }





\end{document}